\newcommand{\CLASSINPUTtoptextmargin}{1.85cm}
\documentclass[a4paper, 9pt, conference]{IEEEtran} 
\IEEEoverridecommandlockouts
\usepackage{cite}
\usepackage{amsmath,amssymb,amsfonts}
\usepackage{graphicx}
\usepackage{float}
\usepackage{textcomp}
\usepackage{xcolor}
\usepackage{subfigure}
\usepackage{multirow}
\usepackage{scalerel}
\usepackage{dblfloatfix} 
\usepackage{algorithm} 
\usepackage{algpseudocode} 

\usepackage[normalem]{ulem}
\usepackage[utf8x]{inputenc}

\begin{document}

\title{Joint Channel Estimation and Synchronization Techniques for Time Interleaved Block Windowed Burst OFDM}


\author{João Martins\IEEEauthorrefmark{1}, Filipe Conceição\IEEEauthorrefmark{1}, Marco Gomes\IEEEauthorrefmark{1}, Vitor Silva\IEEEauthorrefmark{1}, Rui Dinis\IEEEauthorrefmark{2}

\thanks{
}
\IEEEauthorblockA{\\\IEEEauthorrefmark{1}University of Coimbra, Instituto de Telecomunicações,\\ Department of Electrical and Computer Engineering, \mbox{3030-290} Coimbra, Portugal;\\ João Martins (J.M.); filipe.conceicao@co.it.pt (F.C.); marco@co.it.pt (M.G.); vitor@co.it.pt (V.S.)\\}
\IEEEauthorblockA{\IEEEauthorrefmark{2}Universidade Nova de Lisboa (UNL), Instituto de Telecomunicações, \\Faculdade de Ciências e Tecnologia (FCT), \mbox{2829-516} Caparica, Portugal; \\
rdinis@fct.unl.pt (R.D.)\\} 
}

\maketitle

\begin{abstract}

From a conceptual perspective, 5G technology promises to deliver low latency, high data rate and more reliable connections for the next generations of communication systems. To face these demands, modulation schemes based on Orthogonal Frequency Domain Multiplexing (OFDM) can accommodate these requirements for wireless systems. On the other hand, several hybrid OFDM-based systems such as the Time-Interleaved Block Windowed Burst Orthogonal Frequency Division Multiplexing (TIBWB-OFDM) are capable of achieving even better spectral confinement and power efficiency.
This paper addresses to the implementation of the TIBWB-OFDM system in a more realistic and practical wireless link scenarios by addressing the challenges of proper and reliable channel estimation and frame synchronization. We propose to incorporate a preamble formed by optimum correlation training sequences, such as the Zadoff-Chu (ZC) sequences.
The added ZC preamble sequence is used to jointly estimate the frame beginning, through signal correlation strategies and a threshold decision device, and acquire the channel state information (CSI), by employing estimators based on the preamble sequence and the transmitted data. 
The employed receiver estimators show that it is possible to detect the TIBWB-OFDM frame beginning and provide a close BER performance comparatively to the one where the perfect channel is known.

\end{abstract}



In the segment of communications industry, wireless mobile communications are one of the most important innovations of our times due to its impact in the global economy and society. Expectations grow with the arrival and demands that the fifth generation of mobile telecommunications (5G) brings associated with reliable and quality everywhere user access \cite{Vaezi_MultipleAcess5G}.
In this context, several hybrid modulation waveforms, such as the Time-Interleaved Block Windowed Burst Orthogonal Frequency Division Multiplexing (TIBWB-OFDM) \cite{tibwbOFDM, tibwb_ibdfe,tibwb_WTO,Survey5Gwaveforms}, have been recently proposed. Based on the BWB-OFDM \cite{bwbOFDM} modulation scheme, this technique guarantees better spectral efficiency and low power efficiency in comparison with conventional OFDM systems. The great advantage of this approach is that it creates a diversity effect in the frequency domain, making possible to recover part of the lost information caused by frequency selective channel deep fades \cite{tibwbOFDM}.
Additionally, it can be easily implemented for multiple-input multiple-output (MIMO) and massive MIMO systems \cite{tibwb_MIMO}. 
In this paper, we focus on the deployment of this modulation system in a more realistic scenario by addressing to channel estimation and symbol synchronization problems. These are in fact, two important challenges to deal in the practical deployment of any wireless systems since the transceiver radio channel can be highly dynamic, which corrupts the transmitted signal and limits the system performance \cite{OFDM_EstimationOzdemir}. 
Nevertheless, if the receiver is able to, first, detect correctly the frame beginning and afterwards estimate accurately the channel characteristics then, it is possible to recover information reliable.

In order to perform timing synchronization, it is important to first detect the presence of data, by measuring the signal received power, and then estimate the beginning of the preambles sequences \cite{SoftwareDefinedRadio2018}. This can be made by searching for periodic structures within the signal, i.e., applying correlation signal techniques between the received frame and known markers. Therefore, the use of cross-correlation is a possible method to provide an appropriate frame detection \cite{RobustSyncOFDM_Schmidl}. 
For that, it is important to carefully choose optimum preamble sequences for the system acquisition process. In the LTE communication schemes, primary synchronization signals (PSS) are composed by Zadoff-Chu (ZC) sequences \cite{chucodes1972}, complex-valued mathematical sequences with good properties to use in synchronization techniques \cite{Dahlman4G_to_5G}. 

In terms of channel estimation, the most usual approach in OFDM-based systems are either inserting known pilot symbols or training sequences in the transmitted frame \cite{pilotpattern2000}. There are several possibilities to adopt: 

\begin{itemize}
\item The block-type pilot allocation, used in IEEE 802.11a/g/n standard, corresponds to allocate pilots in the frequency domain into all subcarriers. Typically are designed for slow-fading frequency selective channels when the OFDM symbols time duration is much smaller than the channel coherence time \cite{Chiueh2012BasebandMIMO};
\item The comb-type pilot allocation, used, for instance, in the IEEE 802.11a WLAN standard \cite{ManPunMicheleMorelli2007}, corresponds to insert pilots, at predefined subcarriers locations across all entire transmission time with the purpose to resist fast channel time variations between OFDM symbols. However, it is crucial to guarantee the spacing between each pilot subcarriers to be much smaller than the channel coherence bandwidth for effective and accurate estimation;
\end{itemize}
Topologies where pilots are scattered over time and frequency domain can also be taken into account, enabling a good tracking relationship between frequency selectivity and time variation of the wireless channel. Since the pilot's grid insertion is not made in all subcarriers or in fixed subcarriers across all the time, it is possible to achieve a better overall system performance by reducing the pilot density and thus improving the spectral efficiency \cite{pilotpattern2000}. 

In our work, we propose using a ZC preamble sequence to jointly perform channel estimation and frame synchronization which can offer a close performance comparatively to the achieved with perfect CSI. Hence, this paper offers a perspective of the techniques that could perform well in the TIBWB-OFDM system, including the preamble block-type allocation strategy, the frame correlators algorithm and the channel estimators for a linear equalizer receiver structure or by adapting the iterative block decision feedback equalization (IB-DFE) \cite{ibdfe_1,ibdfe_2} structure.

The remainder of the paper is organized as follows. Section \ref{sc:2} gives an overview of all the relevant scientific knowledge required to provide a clearer perspective of the presented work. Section \ref{sc:3} introduces the proposed TIBWB-OFDM’s joint channel estimation and frame synchronization strategy along with a set of adapted iterative equalizers. Section \ref{sc:4} presents the discussion and analysis of the results of channel estimation algorithms and frame detection for the TIBWB-OFDM system. Finally, conclusions and future work proposals are presented in Section \ref{sc:5}.

\section{Background}\label{sc:2}

In this chapter we introduce the background knowledge of the TIBWB-OFDM modulation systems with special attention for symbol equalization, synchronization and channel estimation techniques.

\subsection{TIBWB-OFDM Waveform}\label{sc:2.1}

The building process of the TIBWB-OFDM transmitted block, $\mathbf{X}$, consists into packing together a set of $N_s$ windowed OFDM symbols by a time-domain square root raised cosine (SRRC) window \cite{Haykin_CommunicationSystems} of roll-off $\beta$, followed by a time-interleaving (permutation) operation.

Let $\mathbf{S}_{i} = [S_{i,0} , S_{i,1} , ... , S_{i,(N-1)}]^T$ denote the $i^{th}$ conventional $N$-carrier OFDM symbol packed within $\mathbf{X}$, with $i=0,...,N_{s-1}$, and let $\mathbf{S} = [\mathbf{S}_{0} , \mathbf{S}_{1} , ... , \mathbf{S}_{N_{s-1}}]$.
The construction of block $\mathbf{X}$ can be represented  in matrix format as
\begin{equation}
    \mathbf{X} = \Pi^{\scriptscriptstyle (N_s)} vect\left [\mathbf{A}\left((\mathbf{1}^T_{N_s} \otimes \mathbf{h}_{SRRC} )\odot(\mathbf{1}_{2}\otimes\mathbf{F}\mathbf{S})\right)\right],
\end{equation}
where:
\begin{itemize}
    \item $\mathbf{F}$ is the Inverse Discrete Fourier Transform (IDFT) matrix with size $N{\times} N$ ;
    
    \item $\mathbf{h}_{SRRC} = [h_{-N}, ... ,h_{-1},h_0, ... ,h_{N-1}]^T$ is the SRRC is the SRRC of roll-off $\beta$ where

\begin{equation} \label{equation_srrc}
\mathrm{h}_{n}=\left\{\begin{matrix}
 1 &  ,\left | n \right |\leq  \frac{N}{2}(1-\beta)  \\
cos(\frac{\pi}{4\beta}[\frac{2n}{N}-(1-\beta)]) & ,\frac{N}{2}(1-\beta) \leq  \left | n \right | < \frac{N}{2}(1+\beta)\\ 
0 & ,\left | n \right |\geq \frac{N}{2}(1+\beta)
\end{matrix}\right. \quad
\end{equation}

\item $\mathbf{A}=[\mathbf{0}_{N(1{+}\beta) \times \frac{N}{2}(1{-}\beta)} \quad \mathbf{I}_{N(1{-}\beta)} \quad \mathbf{0}_{N(1{+}\beta) \times \frac{N}{2}(1{-}\beta)}]$ is a truncation matrix, that removes the trailing and ending rows of zeros that result from the windowing operation;

\item $\mathbf{1}_\ell$ is a $\ell$-length column vector of $1's$, $\mathbf{0}_{m{\times}n}$ is  $(m{\times}n)$-size matrix of $0's$ with , and $\mathbf{I}_{n}$ is is  $(n{\times}n)$-size identity matrix;

\item $vect()$ is the matrix vectorization function, that reshapes a matrix into a column, by reading the matrix column by column;

\item $\odot,\otimes,\times$ denote, respectively, the Hadamard, Kronecker and convectional matrix multiplications;

\item $\Pi^{\scriptscriptstyle (N_s)}$ is the time-interleave permutation matrix \cite{tibwb_ibdfe}.
\end{itemize}

Finally, previous to transmission a zero-pad (ZP), with length $N_{ZP}$ greater than the propragation delay, is added at the end of each TIBWB-OFDM block in order to avoid any Inter-Symbol Interference (ISI) \cite{tibwb_WTO}. 

\subsection{Frame Synchronization and Zadoff-Chu Sequences}\label{sc:2.2}

Frame detection based on the use of either pilot or preamble sequences, known at the receiver, is a classic strategy used to achieve synchronization. Typically, these markers have proprieties that facilitate the frame estimation detection \cite{SoftwareDefinedRadio2018}.
Briefly, the frame beginning position is estimated by performing a correlation signal between the received frame, $y$, and the known $L_p$ length preamble, $p_n$ \cite{RobustSyncOFDM_Schmidl}, according to 

\begin{equation} \label{eq:ccorrelation}
\textrm{C}_{y,p_n}(l)= \sum_{m}y(m)p_n^*(l+m).
\end{equation}
where $p_n^*(l)$ is the conjugate of the preamble sequence.

Depending on the chosen preamble sequence and their proprieties, it is expected that when $y \approx p_n$, at least a high peak is produced. 
When performing the auto-correlation, the highest peak, $C_{max}$, is given by
\begin{equation} \label{correlationpeak_max}
\textrm{C}_{max} = \sum_{m} \left | p_n(m) \right |^2,
\end{equation}
where the total correlation length is equal to $2L_p-1$.

When the frame's estimated starting point is obtained, it is important to decide if the marked position is indeed the beginning of the frame or not, i.e., define a detection probability of the following events
\begin{equation}
\left\{\begin{matrix} 
\mathcal{H}_0 : \textrm{Signal absence} \\ \mathcal{H}_1 : \textrm{Signal detected}
\end{matrix}\right. \Leftrightarrow \left\{\begin{matrix} 
\mathcal{H}_0 : y = \eta \\ \mathcal{H}_1 : y = x_n + \eta
\end{matrix}\right.,
\end{equation}
where for the absence hypothesis, the received signal is composed only by the noise component, $\eta$, and, on the other hand, the detected hypothesis is the superposition between the noise, and transmitted signal blocks, $x_n$. 

Based on $C_{max}$ and the noise distribution, it is important to define a threshold, $\delta_{decision}$ and a decision rule such that considering a length-2$L_p$ sliding window analysis of (\ref{eq:ccorrelation}), within each analysis interval $\mathcal{L}$

\begin{equation}
\left\{\begin{matrix} 
\text{select} \ \  \mathcal{H}_1 \quad if \quad \underset{l \in \mathcal{L}}{max} \, C_{y,p_n}(l) > \delta_{decision} \\ \text{select} \  \ \mathcal{H}_0 \quad if \quad \underset{l \in \mathcal{L}}{max} \, C_{y,p_n}(l) < \delta_{decision}
\end{matrix}\right. .
\end{equation}

The $\delta_{decision}$ must be set in order to maximize the success frame detection probability ($P_D$) and reduces situations of either false alarm ($P_F$) or missed frame detection ($P_M$).
Mathematically these probabilities definitions can be expressed as

\begin{equation}
    P_D = P_r \left \{  \mathrm{Select \,\,} \mathcal{H}_1|\mathcal{H}_1\right \}\times P_{r}\left \{H_1\right \} + P_r \left \{  \mathrm{Select \,\,} \mathcal{H}_0|\mathcal{H}_0\right \}\times P_{r}\left \{H_0\right \},
\end{equation}

\begin{equation}
    P_M = P_r \left \{  \mathrm{Select \,\,} \mathcal{H}_1|\mathcal{H}_0\right \},
\end{equation}

\begin{equation}
    P_F = P_r \left \{ \mathrm{Select \,\,} \mathcal{H}_0|\mathcal{H}_1\right \}.
\end{equation}

\paragraph{\textbf{Zadoff-Chu Sequences}}

An essential part for achieving a reliable signal recovery is the choice of optimum preamble sequences.
In the LTE standard, as part of the system acquisition process, two preamble sequences, defined as PSS  and secondary synchronization signals (SSS), are used in the downlink process \cite{KhanLTE2009}. Although these SSS sequences are based on maximum length sequences, also known as M-sequences, the PSS is composed by Zadoff-Chu sequences \cite{chucodes1972} of odd-length period $N_{ZC}$, given by \cite{Dahlman4G_to_5G}, 
\begin{equation}
zc_{n}{[q]}=e^\frac{-j\pi n q (q+1)}{N_{ZC}} , \quad q = 0,1, ... , N_{ZC}-1,
\end{equation}
where \textit{n} is the root index and $n \in \{1, ... , N_{ZC} - 1\}$.

As referred, ZC sequences exhibit some fundamental proprieties that aid the receiver in the synchronization task \cite{SesiaLTE2011,KhanLTE2009}: 

\begin{itemize}
\item They have a constant amplitude, which, on the other hand, limits the Peak-to-average Power Ratio (PAPR);
\item They have "perfect" cyclic auto-correlation, i.e. the correlation with its circularly shifted version is zero at samples different from $N_{ZC}$ and non-zero only at one instant which corresponds to the $N_{ZC}$ sample. The zero auto-correlation ZC property can be formulated by \eqref{eq:ccorrelation}, resulting in

\begin{equation} \label{eq:ccorrelation_chu}
\textrm{C}_{zc,zc}[l]= \sum_{m}zc[m]zc^{*}[m+l] = P_{zc}[l],
\end{equation}
where $P_{zc}$ corresponds to the auto-correlation peak positioned in the delayed l sample. This is an important feature for wireless communications enabling misaligned ZC sequences to correlate between themselves. Hence, it is possible to generate multiple orthogonal sequences just by shifting the ZC sequence.

\item If we guarantee the Zadoff-Chu sequence length $N_{ZC}$ to be a prime number, then, the cross-correlation of these two sequences is $1/\sqrt{N_{ZC}}$.
\end{itemize}

Sequences that combine the first two properties are designated as Constant Amplitude Zero Auto-Correlation (CAZAC) sequences \cite{SesiaLTE2011}.
A crucial application of ZC sequence is for time synchronization due to their CAZAC properties. Therefore, the receiver by analysing the sequence correlation peak inside a timing slot can easily identify the transmitted frame beginning \cite{Hua2014_ZCAnalysis}.

However, ZC sequences are affected by the channel fading delay effects resulting in additional time-domain correlation peaks.
Furthermore, the sequence correct correlation peak amplitude is attenuated at high Doppler frequencies, which can lead to false detect or miss the frame beginning from the several time delay correlation peaks, causing to miss calculate the channel total delay and, consequently, decreasing the overall system performance \cite{Dahlman4G_to_5G}. 

\subsection{Equalization and Channel Estimation}\label{sc:2.3}

The receiver key task is to perform equalization and unformat the received block. The latter operation is established by executing the opposite of the transmitter scheme, i.e, perform the block time-deinterleaving, matched filtering, demodulation, and channel decoding.
The received TIBWB-OFDM block is converted to the frequency domain through a  discrete Fourier time operation (DFT) and assuming that the used ZP is long enough to avoid ISI, frequency domain equalization (FDE) is executed: either linear FDE such as zero-forcing (ZF) or minimum mean square error (MMSE) methods or non-linear iterative equalization block based methods, such as the IB-DFE \cite{tibwb_ibdfe} may be employed at this stage.
When it comes to the FDE of the TIBWB-OFDM symbol, this has several advantages over time-domain equalization (TDE) in outdoor high-mobility propagation environments and for channels with severe delay spreads, since the receiver complexity can be kept low \cite{new_directions_tarokh}.
In channels whose impulse responses remain constant within one transmitted symbol period, the frequency domain received TIBWB-OFDM signal, $Y_k$, at each subcarrier $k = 0, ... , N_{b}$ (with $N_{b} = N_{s}\times N(1+\beta) - 1$) can be expressed as
\begin{equation} \label{channel_model}
    Y_k =H_k X_k+N_k,
\end{equation}
where $H_k$ represents the Channel Frequency Response (CFR). By employing linear FDE the estimated signal is given by
\begin{equation} \label{f_response}
    \hat{X}_k =F_k Y_k,
\end{equation}
where $F_k$ is the frequency response of the feedforward equalization filter.

The two most popular linear equalization schemes are the ZF and the MMSE equalizer.
The ZF equalizer simply uses the inverse of the CFR, i.e,
\begin{equation} \label{Fk}
    F_k = \frac{H_{k}^*}{|H_k|^2} = \frac{1}{H_k},
\end{equation}
with the received signal being given by
\begin{equation}
    \hat{X}_k = \frac{H_k X_k}{H_k} + \frac{N_k}{H_k} = X_k + \frac{N_k}{H_k}.
\end{equation}

Although its computational simplicity, this equalization technique results in noise enhancement caused on the term $\frac{N_k}{H_k}$ especially in the carriers that suffer deep fading. 

In the case of an MMSE receiver, it tries to minimize $E\{|\hat{X}_k - X_k|^{2}\}$, by taking the SNR component, $\gamma$, into account. Therefore, the equalization weight on the subcarrier k is given as
\begin{equation}
    F_k = \frac{H_{k}^*}{|H_k|^2 + \frac{1}{\gamma}},
\end{equation}
resulting in
\begin{equation} \label{Xk_eq}
    \hat{X}_k = \frac{H_{k}^*H_k X_k}{|H_k|^2 + \frac{1}{\gamma}} + \frac{N_k H_{k}^*}{|H_k|^2 + \frac{1}{\gamma}} = \frac{|H_k|^2 X_k}{|H_k|^2 + \frac{1}{\gamma}} + \frac{N_k H_{k}^*}{|H_k|^2 + \frac{1}{\gamma}} .
\end{equation}

This equalizer has the advantage that it minimizes the noise enhancement problem in low SNR regimes, although it does not make a perfect inversion of the channel. Also, when the SNR is high enough, it is clear from \ref{Xk_eq} that the MMSE equalizer approaches the zero forcing equalizer \cite{Chiueh2012BasebandMIMO}.

However, the performance is far from the matched filter bound (MFB) so the alternative approach arrives by adopting nonlinear schemes, such as, decision feedback equalizers \cite{Souto2015Interference}. 
The IB-DFE equalizer is particularly suited for block-based single carrier (SC) transmissions and can deal with ISI and Inter-block Interference (IBI) \cite{tibwb_ibdfe}, by means of both feedforward filter (${F}_{k}^{l}$), that acts like a conventional FDE in order to decrease the precursors created by the wireless channel, and the feedback filter (${B}_{k}^{l}$) that attempts to cancel the remaining interferences through information from previous precursors.

Note that in the case of a TIBWB-OFDM transmission the whole received block is first processed as being of the type of a block-based SC transmission \cite{tibwbOFDM}. By employing IB-DFE the equalized TIBWB-OFDM symbol is given by
\begin{equation}
\tilde{X}_k^{(l)} = F_{k}^{(l)}Y_{k} - B_{k}^{(l)}\hat{X}_{k}^{(l-1)},
\end{equation}
and $\hat{{X}}_{k}^{(l-1)}$ denotes the frequency domain signal estimated at the previous iteration, by either implementing a $\mathit{hard}$ or $\mathit{soft}$ feedback filter decision \cite{tibwb_ibdfe}. 
The optimal feedforward and feedback filter coefficients are, respectively, given by
\begin{equation} \label{eq:FF_filter}
   F_k^l = \frac{\kappa H_k^*}{\frac{1}{\gamma} + \left(1-(\rho_{blk}^{l-1})^2\right)|H_k|^2},
\end{equation}
and 
\begin{equation} \label{eq:FB_filter}
   B_k^l = \rho_{blk}^{l-1}\left(F_k^lH_k - 1\right),
\end{equation}
where $\kappa$ denotes a normalized constant chosen to guarantee that $\frac{1}{N}\sum_{k=0}^{N-1}F_k^lH_k=1$ and $\rho_{blk}$ represents the correlation factor, a key parameter for a reliable IB-DFE operation and a accurate system performance \cite{tibwb_ibdfe,ibdfe_2}.


A challenging obstacle in any communication system for the success of the equalization procedure is the demanded necessity for the receiver to obtain the CSI, and getting frequently updates on it which is not an easy task due to time variance and frequency selectivity of wireless channels.
The pilot sequence length and optimization should therefore be optimized considering the CIR length with the objective of reducing channel estimation errors, the BER and maximizing channel capacity \cite{Ohno2011PreambleSubcarriers}.
Channel estimation techniques can be classified into two categories \cite{Fazel2008Multi-CarrierWiMAX,Chiueh2012BasebandMIMO}:

\begin{itemize}
\item Non-data-aided, where the CSI is obtained without the use of reference training signals, i.e., based on the statistics of the received signals sequences;
\item Data-aided, which require added information, i.e., reference training signals that are included in the transmitted frame.
\end{itemize}

Although non-data-aided techniques do not require reference signals, a large number of data must be collected in order to obtain reliable estimation channel samples. Hence, data-aided techniques, although requiring an additional frame overhead on the transmitted data, that causes a decrease in the spectral efficiency, can provide better performance, especially on fast changing channel conditions.

The simplest method for frequency channel estimation is to use block-type allocation and employ the least-square (LS) estimation, $\hat{H}_{LS}$, wherein each $k^{th}$ subcarrier it is necessary to compute \cite{OFDM_EstimationOzdemir}

\begin{equation} \label{eq:h_ls1}
    \hat{H}_{LS_{k}} = \frac{Y_{k}}{X_{k}} = H_{k} + \frac{N_{k}}{X_{k}}.
\end{equation}

Clearly, the LS estimator constitutes a low complex algorithm. However, the major disadvantage comes from noise enhancement. Since LS estimators do not generally require any channel statistics, the estimation is not perfect enough, especially in scenarios of very fast channel variations where the system performance will significantly deteriorate \cite{Wang2013ChannelStandard}.
Nevertheless, it is an essential task to obtain a first initial coarse channel estimation.

\section{Synchronization and Channel Estimation for TIBWB-OFDM}\label{sc:3}

In order to perform signal synchronization and channel estimation tasks for TIBWB-OFDM, both transmitter and receiver schemes presented in \cite{tibwbOFDM} need to be improved for a practical deployment.
The following two sub-sections explain with more detail the new proposed transceiver architecture.

\begin{figure}[H]
\includegraphics[width=\linewidth]{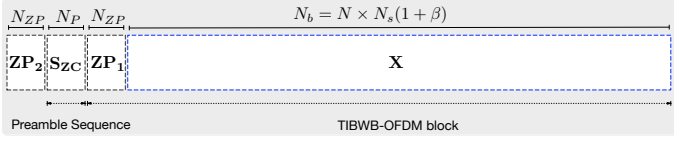}
\caption{TIBWB-OFDM block frame structure adopted.}
\label{fig_tx_tibwb}
\end{figure}

The inherent characteristics of the TIBWB-OFDM scheme favors the adoption of a preamble allocation strategy due to the packing of several OFDM symbols into a single block.
So, as visible in Figure \ref{fig_tx_tibwb}, the interleaved block vector is then attached to the $N_{p}$ length preamble sequence, $S_{ZC}$, forming a new TIBWB-OFDM frame-block.

Due to their auto-correlation properties and low PAPR, ZC sequences already employed in LTE are a good candidate. By adding an additional ZP, the same sequence can be used for CSI estimation. Note that since a single ZC sequence is sent per several OFDM blocks packed within the TIBWB-OFDM block, the spectral efficiency suffer a minimum decrease.

Settling the transmitter block configuration, it is now vital to develop the receiver architecture.

\begin{figure}[H]
\includegraphics[width=\linewidth]{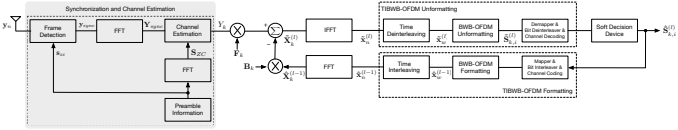}
\caption{Preamble Assisted TIBWB-OFDM receiver scheme with IB-DFE equalization.}
\label{fig:tibwb_rx}
\end{figure}

Figure \ref{fig:tibwb_rx} shows an additional new sub-system, named as \textit{Synchronization and Channel Estimators}, essential to accomplish signal recovery, at a first instance, and the frequency domain channel estimation and equalization. 
On the other hand, it is necessary to know precisely the preamble transmitted sequence. 
Thus, the stored ZC preamble sequences constitutes a key element of the receiver structure.
The respective receiver scheme shows the equalizer implementation specifically for the nonlinear IB-DFE type. If the desired receiver structure is a simple linear equalizer, then, only the TIBWB-OFDM unformatting is performed, obtaining the final estimated sequence, as a one iteration transceiver system.
Afterwards, the TIBWB-OFDM structure is composed by equalizers and block unformatting, i.e., block time de-interleaving, bit-deinterleaving, de-mapping, and channel decoding.

\subsection{Frame Synchronization}

\begin{figure}[H]
 \includegraphics[width=\linewidth]{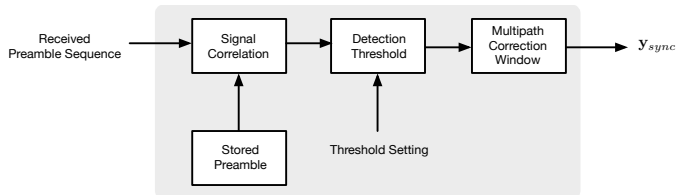}
\caption{Frame detection algorithm procedure for the ZC preamble in the TIBWB-OFDM receiver system.}
\label{fig:algorithm_step}
\end{figure}

Taking  advantage from the ZC proprieties, a time-domain  correlation operation can be performed between the preamble received signal and the known sequence in order to obtain the beginning of each transmitted frame. 
Hence, by auto-correlating the known preamble sequence the maximum correlation peak, $P_{zc}$, is expected to appear in the $N_{ZC}$ sample. This value is extremely important since it will define the threshold percentage value for the receiver system markers and, therefore, affects the success/miss probability of detection.
Figure \ref{fig:algorithm_step} shows the frame detection algorithm flow graph.

In order to better understand Figure \ref{fig:algorithm_step}, Algorithm 1  details in pseudo-code the procedure follow for frame detection. The input variables are the TIBWB-OFDM modulated block, $\mathbf{X}$, composed by a Zadoff-Chu preamble sequence, $\mathbf{S_{ZC}}$, and the threshold setting parameter, $\delta_{decision}$. For the output result it is expected to obtain the correlation indexes peaks of several transmitted preamble frames and the stored ZC sequence, $y_{sync}$, that translate the frame beginning sequence;

\begin{algorithm}
	\caption{Frame Detection} 
	\begin{algorithmic}[1]
		\State Input: $\mathbf{X}$, $\mathbf{S_{ZC}}$ and $\delta_{decision}$.
		\State Output: $y_{sync}$;
		\State $y_{auto\_corr} \leftarrow$ Perform auto-correlation of $\mathbf{S_{ZC}}$;
		\State $y_{corr} \leftarrow$ Perform cross-correlation between $\mathbf{S_{ZC}}$ and $\mathbf{y}$;
		\State $threshold \leftarrow \delta_{decision} \times \mathbf{max}\{|y_{auto\_corr}|\}$;
		\State $y_{thold} \leftarrow y_{corr} \geq threshold$;
		\State $y_{index} \leftarrow$ Find indexes peaks in $y_{thold}$
		
		\For {$iteration=1:2L_{p}$}
		\State $y_{sync} \leftarrow$ Tracking peaks in $y_{index}$ within an $\mathcal{L}$ sliding window and stores the higher peak value;
		\EndFor
	\end{algorithmic} 
\end{algorithm}

The threshold decision device implementation is not straightforward due to the wireless channel multipath effect. 
So, to achieve a good performance, the threshold variable, $\mathit{threshold}$, is not fixed and fluctuates depending on the transmitted correlation preamble amplitude peak.
Also, more than one peak may appear in dispersive channels situations due to signal echos. Therefore, after the detection threshold signal, $y_{thold}$, it is necessary to develop a correction algorithm based on a moving window width, $\mathcal{L}$, and the marked threshold indexes, $\mathit{y_{index}}$, that accepts the maximum peak index and eliminates the remaining ones inside the respective window.
The unique amplitude peak marks the presence of the frame TIBWB-OFDM beginning. However, if we would encounter other high amplitude peaks where should not be expected, the algorithm acknowledges this event has a possible noise frame, rejecting the respective block.

\subsection{Channel Estimation}
After the frame synchronization procedure, the CSI is acquired. Three algorithms were developed for estimating the CSI:
\begin{itemize}
\item ZC preamble based Channel Estimation (Algorithm A);
\item Data based Channel Estimation (Algorithm B);
\item Combined ZC and Data Channel Estimation (Algorithm C).
\end{itemize}

\paragraph{\textbf{Algorithm A}}
Having knowledge of estimated $y_{sync}$, received samples corresponding to $[ZP,ZC]$ are used to estimate the channel. An $FFT$ of length $N_{ZP}+N_p$ is applied to the received samples and divided by the corresponding transmitted samples, i.e., a LS estimator is used according to

\begin{equation}
    \widetilde{H}_{k(zc),(N_{ZP}+N_p)} = \frac{FFT_{N_{ZP}+N_p}\{\widetilde{s}_{zc}\}}{FFT_{N_{ZP}+N_p}\{s_{zc}\}},
\end{equation}
where $\widetilde{s}_{zc}$ represents the received preamble sequence. We admitted that the $N_{ZP}$ sequence length is always larger than the CIR. Thus, enabling the reception of $\widetilde{s}_{ZC}$ without ISI.
The low resolution (in frequency) estimated CSI, is then upconverted by interpolation to the length of the TIBWB-OFDM block

\begin{equation}
    \widetilde{H}_{k(zc),(N_b)} = FFT_{(N_b)} \left \{ IFFT_{N_{ZP}+N_{p}} \left \{\widetilde{H}_{k(zc),(N_{ZP}+N_{p})}\right \}  \right \},
\end{equation}
where $\widetilde{H}_{k(zc),(N_b)}$ represents the estimated channel constituted by $N_b$ samples.
\paragraph{\textbf{Algorithm B}}

If the IB-DFE is adopted it is possible to take advantage of the estimated data $\widehat{X}_k^{(l-1)}$, at iteration $l-1$, to obtain an estimation of the channel to be used at iteration $l$. This is given by

\begin{equation}
    \widetilde{H}_{k(data)} = \frac{Y_k}{\widehat{X}_k^{(l-1)}},
\end{equation}
where $\widehat{X}_k^{(l-1)}$ denotes the frequency domain signal estimated at the previous $(l-1)$ iteration with $N_b$ samples.
\paragraph{\textbf{Algorithm C}}

A refined strategy for achieving successive better estimations can be used leveraging on estimations from algorithm B.
Note that once CSI is obtained with Algorithm A based on the known preamble sequence it makes no sense to disregard such information and consider only Algorithm B based on estimated data. Instead Algorithm C consider a weighted mean between the ZC channel estimated sequence (algorithm A) and the data sequence estimated sequence (algorithm B) can be applied as follows

\begin{equation}
\widetilde{H}_{k(data+zc)}=\frac{\left ( \frac{\widetilde{H}_{k(data)}}{\sigma_{(data)}^{2}} + \frac{\widetilde{H}_{k(zc)}}{\sigma_{(zc)}^{2}} \right )}{\left (\frac{1}{\sigma_{(data)}^{2}} +  \frac{1}{\sigma_{(zc)}^{2}} \right )},
\end{equation}
where $\sigma_{(zc)}^{2}$ and $\sigma_{(data)}^{2}$ represents individually the mean square error between the estimated ZC and data sequence, respectively, and the original channel. This relation is given by
\begin{equation}
\sigma_{(zc / data)}^{2} = \textbf{E} \left \{ \left | H_k - \widetilde{H}_{k_{(zc / data)}} \right |^2 \right \}.
\end{equation}

\section{Experimental Evaluation}\label{sc:4}

To test the behavior of the algorithms two severe Rayleigh channel scenarios were considered, respectively with 8 and 32 symbol-spaced multipath channels echos (channel taps) with uncorrelated Rayleigh fading were used to evaluate the frame detection probability error and the channel estimators performance through analysis of the BER of the received signal. 
The parameters for the TIBWB-OFDM wireless SISO model simulations are defined in table \ref{table:1}.

\begin{table}[H]
\caption{Simulation Parameters}
\begin{tabular}{ cccc } 
\hline
\textbf{Parameter}  & \textbf{TIBWB-OFDM}  & \textbf{Preamble}  \\
\hline
Shaping Pulse & SRRC & - \\ 
Roll-off factor & $\beta$ = 0.5 & - \\
Modulation & QPSK & - \\ 
OFDM Symbols & $N_s = 42$ & - \\
Root Element & - & n = 34 \\ 
Sub-Block Length & $N = 64$ & $N_{ZC} = 95$ \\
Block Size & $N_{b} = 4032$ & $N_{p} = 96$\\
LDPC & (128,64) & - \\
Bit-Interleaving & 10 consecutive words & - \\
\hline
\end{tabular}
\label{table:1}
\end{table}

Since it was expected that successful frame synchronization would be dependent on the power of the sent ZC preamble, we also consider, three cases where the average power of the transmitted ZC sequence was the same, 3 or 6 dB higher than the average TIBWB-OFDM data power block. Also, we wanted to evaluate the impact of this feature on the channel estimators' performance.
\subsection{Threshold Awareness Results}

The following simulations aim to evaluate and acquire the best threshold index number that reduces the miss frame probability. Therefore, 700 data frames, i.e, TIBWB-OFDM blocks, are randomly interleaved with 300 equally size blocks of additive white gaussian noise (AWGN) with independent uncorrelated samples.
It is important to enhance that all the preamble signals were identical which reflects on the same transmitted high correlation peak amplitude.  

As referred, one of the critical performance variables is the selection of the threshold cross-correlation level used to indicate the beginning of a new frame. 
Hence, the used value should be as accurate as possible to achieve an acceptable frame detection that will subsequently influence the BER system performance and error awareness. 
\begin{figure}[ht]
 \includegraphics[width=\linewidth]{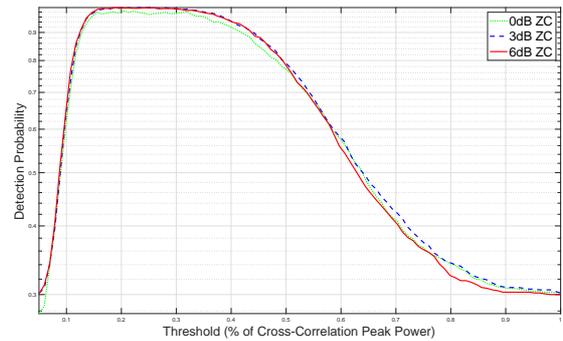}
 \caption{Correct frame detection probability as function of the cross correlation’s threshold value for peak detection.}
 \label{threshold_probability}
\end{figure}

Figure \ref{threshold_probability} shows the probability of correct detection as a function of the setup threshold level. This value should be selected in the range of $25\%-30\%$ of the maximum of the auto-correlation, $P_{zc}$, in which it can be observed that the probability of correct detection is almost perfect.
When the threshold parameter is near $P_{zc}$ no data frames are detected. This is due to the fact that the signal suffers from fading (i.e., variable gain attenuation due to multipath effect), with the cross-correlation value hardly reaching the $P_{zc}$ value. In this case, since almost all frames are marked as noise, the probability of right detection is $30\%$ corresponding to the fraction of noise frames sent.
On the contrary, by settling the threshold too low, the opposite effect occurs with almost all frames being detected as data.
Here, the coexisting noise peaks combined with those of the data sequences, both above the threshold make it impossible to detect a single peak value causing the frame recognition task very challenging to accomplish. 
However, in this scenario, a noteworthy difference is visible between the non-amplified ZC sequences and the remaining sequences. 
Since the threshold is defined as a percentage of $P_{zc}$, for the 3 dB and 6 dB cases, the threshold is much higher precluding any noise frame to be marked as data, and, thus, the probability of a correct detection is above $30\%$.

\subsection{Channel Estimators with Linear Equalization}

After addressing the frame synchronization problem, we will evaluate the effectiveness of the proposed CSI estimators.
Therefore, the subsequent simulations evaluate the BER results of the TIBWB-OFDM modulation scheme in a set condition without the perfect channel state knowledge and compare them with the case of perfect CSI.

\begin{figure}[H]
 \includegraphics[width=\linewidth]{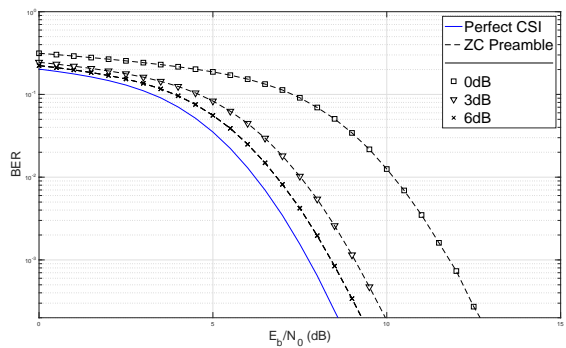}
 \caption{BER results for TIBWB-OFDM over a dispersive channel employing algorithm A for a ZC preamble sequence with power amplification.}
 \label{ber_chu}
\end{figure}

Figure \ref{ber_chu} compares the TIBWB-OFDM BER performances with the perfect CSI and ZC sequences as preambles in order to perform channel estimation. 
For a low computational scenario, algorithm A was chosen as the first estimation approach where the goal was to find how the ZC sequences would perform in channel estimation for TIBWB-OFDM modulation with MMSE equalization.

Although the optimal channel state information is not possible, a good performance approximation comes with the price of a necessary increase power of the preamble sequence. Having as reference the average power of the TIBWB-OFDM data block, we tested 3 situations, where the appended ZC preamble was sent with the same power as data (i.e. 0 dB case), or 3 dB and 6 dB higher.
The system performance shows an evident improvement with power amplification of plus 3 dB. Also, for the 6 dB amplification, it is possible to achieve a slight improvement, almost reaching the reference performance (assuming perfect synchronization and channel estimation).

It is essential to highlight that although it is wasted some additional power for good system performance, this only applies to the preamble of the sub-block, which reflects in an approximate 2\% of additional wasted total system power in the 3 dB scenario, and 5\% in the 6 dB scenario for the considered TIBWB-OFDM frame structure. Hence, it can be concluded that it is not critical to increase our preamble power since the total overall efficiency power calculation is weighted on the data frame and not on the preamble sequence.

\subsection{Channel Estimators with Iterative Frequency Domain Equalization} 

\begin{figure}[H]
 \includegraphics[width=\linewidth]{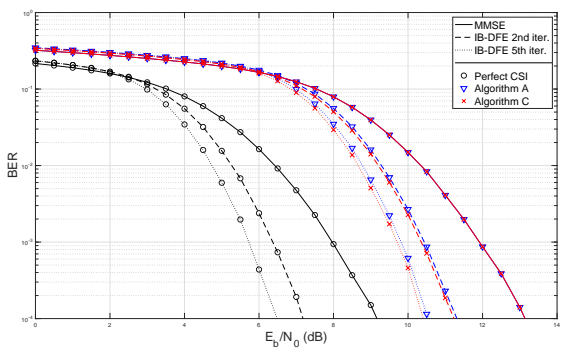}
 \caption{BER results for TIBWB-OFDM with IB-DFE over a dispersive channel employing algorithm \textbf{C} for a ZC preamble having the same average power as data symbols.}
	\label{ibdfe_0dB}
\end{figure}

\begin{figure}[H]
 \includegraphics[width=\linewidth]{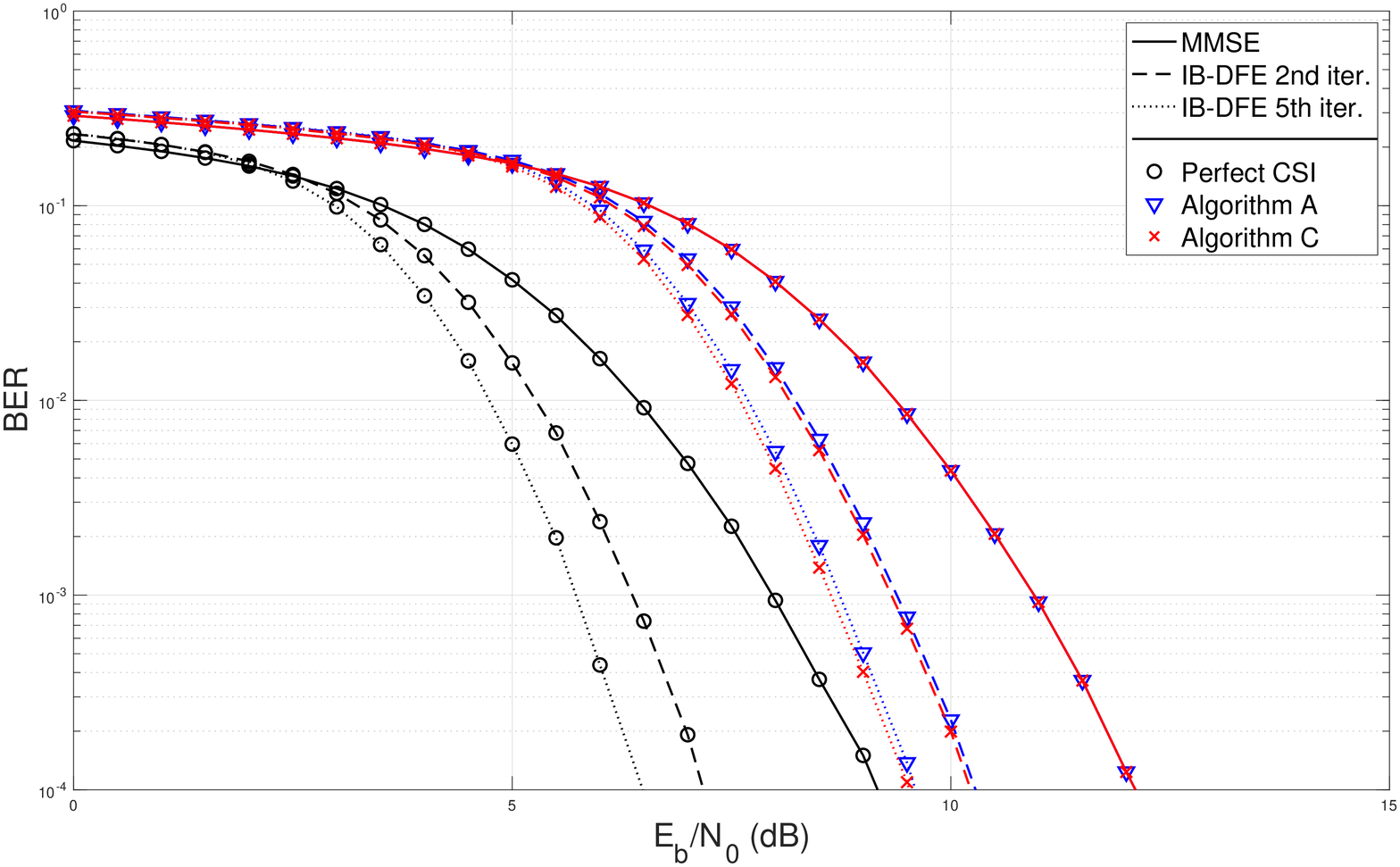}
 \caption{BER results for TIBWB-OFDM with IB-DFE over a dispersive channel employing algorithm \textbf{C} for a ZC preamble sequence with 3 dB higher average power than the data symbols.}
    \label{ibdfe_3dB}
\end{figure}

\begin{figure}[H]
 \includegraphics[width=\linewidth]{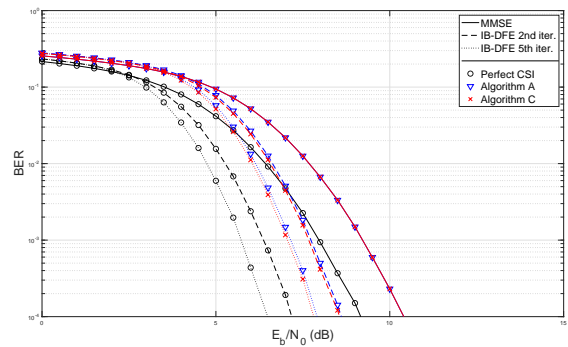}
 \caption{BER results for TIBWB-OFDM with IB-DFE over a dispersive channel employing algorithm \textbf{C} for a ZC preamble sequence with 6 dB higher average power than the data symbols.}
 \label{ibdfe_6dB}
\end{figure}

Aiming to achieve an even better estimation of the TIBWB-OFDM original bit-stream we consider the use of the IB-DFE equalizer and CSI obtained with Algorithm C applied with results being presented in Figures \ref{ibdfe_0dB}, \ref{ibdfe_3dB} and \ref{ibdfe_6dB}, for different values for the preamble average power.

It should be reasonable to think that for low SNR values, the channel would maintain massive influence in the mean square error calculation, and for high SNR values, the opposite effect should happen.
It should also be expected to achieve, at each IB-DFE iteration, a gradual better performance.
However, the results show that even for the worst case (ZC sequence with the same average power of the data is used) the fine CSI estimation does not significantly affect the overall system performance. This can be justified by calling the good performance of ZC on dispersive channels and how ineffective the interference affect the preamble sequence. Also, the data does not present these good characteristics for channel estimation. So, the algorithm's estimated weight is enhanced by the first estimation, performed by the ZC preamble, i.e., by algorithm A. 

Thus, although a slight performance increase is achieved, the algorithm C complexity does not justify this extra estimation procedure. However, it is clear that the IB-DFE receiver shows some performance evolution form iteration to iteration, and also the influence of increasing the power of the preamble. In fact, Figure \ref{ibdfe_6dB} show the effectiveness of Algorithm A combined with IB-DFE, since with just two iterations, it can perform better than the MMSE with perfect CSI, and a further 1 dB gain can be obtained if 5 iterations are considered.

\section{Conclusions}\label{sc:5}

The developed work constitutes a new step to provide an insight on channel estimation and frame detection algorithms that could perform well in disperse channel conditions for the TIBWB-OFDM scheme. Our work shows that the developed estimators by means of block-type preamble assisted frames can provide a close BER system performance in comparison with the true channel state information performance due to the careful choice of the preamble sequences. So, it is useful to implement the pilot design based on ZC sequences due to their good correlation characteristics, and is proven that they are able to jointly execute channel and synchronization estimation effectively in our hybrid modulation architecture. 
Nevertheless, it is necessary to guarantee that the preamble power block is amplified.

Therefore, a proof of concept of how the studied modulation scheme should perform in a real world environment is established. 

Considering the obtained results, future developments should involve the study of this estimators applied to the TIBWB-OFDM with the windowing time overlapping \cite{tibwb_WTO}, and the implementation of the proposed synchronization and channel estimators in a practical real environment, i.e., using Software Defined Radio (SDR) based testbed.

\nocite{*}
\bibliography{IEEEabrv,bibliography}
\bibliographystyle{IEEEtran}

\end{document}